# Prime Factoring and The Complexity Of

Charles Sauerbier

## 1 Introduction

Prime factorization is a mathematical problem with a long history. One of the oldest known methods of factoring is the Sieve of Eratosthenes. There have been numerous methods[1] developed since the time of Eratosthenes. The Pollard Rho method[2] is commonly accepted as the fastest publicly available factoring method.

Prime factoring is important for multiple reasons. One is its central role in several cryptographic schemes. The central element of factoring is to find two prime numbers $p, q$ where $n = p * q$, given the value of $n$. A mathematical expression is developed that is amenable to iterative resolution of prime factors for a composite. The expression is then shown to admit a factoring process of P-complexity. The method may be used to determine if an integer is a prime.

The path to derivation of the expression was an exercise in empirical observation and reasoning a mathematical relation from what was observed, as opposed to any conventional approach of theorizing a means on basis of prior knowledge or approach used by others. As such the paper takes a less than conventional approach to presentation of the material. It is assumed the reader has familiarity with the underlying basic mathematics. Example evaluations are provided for assistance in visualizing operative behavior of expressions.

The derivation flows from well-established mathematical principles, so no proofs are provided. The computational complexity of the resulting expression is a simple exercise in inductive reasoning on linear combinations.

## 2 Basic Concepts

It is a mathematical property of any two factors $p, q : n = p * q$ that one factor is greater than the square root of $n$ and the other less than the square root. Stated as a rule:

[2.1] $\quad n = p * q \leftrightarrow 1 \leq p \leq \sqrt{n} \leq q \leq n$

A second mathematical property of any two factors $p, q : n = p * q$ is that the $\log_2$ of the factors will always be equidistant from the $\log_2$ of the square root of $n$. Stated as a rule:

[2.2] $\quad n = p * q \rightarrow \exists a \geq 0 : (\log_2 p + a) = \log_2 \sqrt{n} = (\log_2 q - a)$

---

[1] See [2].
[2] See [1].

[2.2] admits the use of an iterative search for $p$, $q$. However, it requires real number valued computations using logarithms. The process can be computationally intensive, time consuming to perform and is influenced by numerical precision. Given [2.2] it is not unreasonable to expect a similar relationship exist in the integer field that does not require use of logarithms.

Given some $n = p * q$, if it is presumed the factors have the relationship $p \leq q$ and that $p$ is a prime number the rule [2.3], below, holds, where $m \leq \lfloor \sqrt{n} \rfloor$ and $1 \equiv m \pmod{2}$.

[2.3] $$n = p * q \rightarrow \begin{cases} p = 2, & 0 \equiv n \pmod{2} \\ p = m - 2x, & 1 \equiv n \pmod{2} \end{cases}$$

Given [2.2] and [2.3] it follows that the rule [2.4], below, respective $q$ holds, subject to the constraints on $m$ for [2.3].

[2.4] $$n = p * q \rightarrow \begin{cases} q = n/2, & 0 \equiv n \pmod{2} \\ q = m + 2y, & 1 \equiv n \pmod{2} \end{cases}$$

Given [2.3] and [2.4] basic algebra implies that there exist, for any case where $n$ is odd, a tuple $\langle m, x, y \rangle$ determining the factors $p, q$, and; the factors are described by three equations in four unknowns: $p, q, x, y$.

## 2.1 Difference Expressions

### 2.1.1 Of Ordered Pair $\langle x, y \rangle$

On basis of empirical observation of [2.3] and [2.4] the problem can also be restated as the difference expression [2.5], below.

[2.5] 
$$\langle x_k, y_k \rangle = \begin{cases} \langle x_{k-1} + \varepsilon_x, y_{k-1} \rangle, & \delta_{k-1} < 0 \\ \langle x_{k-1}, y_{k-1} + \varepsilon_y \rangle, & \delta_{k-1} > 0 \\ \langle x_{k-1}, y_{k-1} \rangle, & \delta_{k-1} = 0 \end{cases}$$

$$\delta_0 = \lfloor \sqrt{n} \rfloor^2 - n$$

$$\delta_k = \delta_{k-1} + (\varepsilon_x * x_k) + (\varepsilon_y * y_k)$$

$$\varepsilon_x = -2$$

$$\varepsilon_y = 2$$

The difference expression of [2.5] alternately increments the elements of the ordered pair $\langle x, y \rangle$. If $n$ is an odd composite [2.5] will converge to an ordered pair satisfying [2.3] and [2.4] for two factors $p, q$ such that $n = p * q$.

*Figure 1* presents an example evaluation of [2.5]. It demonstrates an empirically observed pattern in evaluation of [2.5]: That over continuous discrete intervals of $k$ the value of $y_k$ is

     

strictly monotonically increasing, with the value of $x_k$ remaining constant. The pattern results in $y$ forming a discrete approximation to a curve. The points at which $x$ is incremented determining inflection points in an iterated discrete approximation of the curve.

By modifying the computation of [2.5] to set the value of $\varepsilon_y$ to a multiple of 2 sufficient to cause $\delta_k \leq 0$, the number of iterative steps is reduced, as illustrated in example of _Figure 2_.

### 2.1.2  Of Parameters $(\varepsilon_y, \delta_x, \delta_y)$

[2.6] presents the difference expression reformulated in terms of the parameters $(\varepsilon_y, \delta_x, \delta_y)$, where $\varepsilon_x$ is held to a constant -2. _Figure 3_ presents an example evaluation of [2.6]. As observable in _Figure 3_, unlike [2.5], the evaluation of [2.6] halts on $\delta_{y_k} = 0$. The factors of $n$ in this case are $p = x_k$, $q = y_{k+1}$, requiring computation of $y_{k+1}$ to obtain $q$.

[2.6]
$$\varepsilon_0 = 2$$
$$\varepsilon_k = \begin{cases} \left\lVert \frac{\delta_k}{x_k} \right\rVert + 1, & \left\lVert \frac{\delta_k}{x_k} \right\rVert \ (mod\ 2) \neq 0 \\ \left\lVert \frac{\delta_k}{x_k} \right\rVert, & \left\lVert \frac{\delta_k}{x_k} \right\rVert \ (mod\ 2) = 0 \end{cases}$$
$$\delta_{x_0} = \lfloor \sqrt{n} \rfloor^2 - n$$
$$\delta_{x_k} = \delta_{y_{k-1}} + (-2 y_k)$$
$$\delta_{y_k} = \delta_{x_k} + (\varepsilon_k x_k)$$
$$x_k = x_{k-1} - 2$$
$$y_k = y_{k-1} + \varepsilon_{k-1}$$

### 2.1.3  Closed Form Difference Expression

From [2.6] by less than obvious means the closed form expression of [2.7] is obtained, where the values $\langle A \rangle = \{ a_0, a_1, \ldots, a_j \}$ represent the number of occurrences of epsilon values (i.e., $2, 4, 6, \ldots, 2j$ ) in computation of [2.6]. The closed form expression of example of _Figure 3_ is presented in [2.8].

[2.7]
$$x_k = x_0 - (2k + 2)$$
$$y_k = y_0 + \sum_{i=0}^{j}(2i + 2) a_i$$
$$k = \sum_{i=0}^{j} a_i$$
$$\delta_k = n - (x_k * y_k)$$

   

[2.8]
$$A = \{6, 6, 4, 1\}$$
$$\begin{aligned}k &= \sum_{i=0}^{j} a_i \\ &= 6 + 6 + 4 + 1 \\ &= 17\end{aligned}$$
$$\begin{aligned}x_k &= 63 - (2 * 17) + 2 \\ &= 63 - 34 + 2 \\ &= 29 + 2 \\ &= 31\end{aligned}$$
$$\begin{aligned}y_k &= 63 + \sum_{i=0}^{j}(2i + 2)a_i \\ &= 63 + (2 * 6) + (4 * 6) + (6 * 4) + (8 * 1) \\ &= 63 + 68 \\ &= 131\end{aligned}$$

## 3 Factoring

The closed form expression of [2.7] can be resolved by any method that produces appropriate values for $\langle A \rangle$ so that $\delta_k$ attains 0 at some $k$. It admits the use of a binary search in the interval $[1, x_0]$, as means to determine values of $\langle A \rangle$, which limits the maximum number of inflection points at $j \leq (x_0 - 1)/2$.

Factoring using [2.7] can be considered in terms of a *Bin Packing* (BP) problem. The central issue being to determine how many of each $(2i + 2)$ items can be packed into $(\lfloor \sqrt{n} \rfloor - 1)/2$ or fewer bins, such that the items in the bins solve the equality $n = x_k * y_k$. Unfortunately, BP is a problem for which there is not presently any publicly known P-complexity solution for the general case. However, the constraints of the specific case make it a viable conceptual basis for solving the factoring problem.

### 3.1 Limits on Complexity

Considered as a case of the Bin Packing problem there are a finite set of bins defined by $(2i + 2) : i = 0, 1, 2, \ldots, j$ where $j \leq (x_0 - 1)/2$. The objective then is to determine which bin and how many of each are needed to resolve an instance; which leads to the question of whether the number of bins has a tractable limit short of brute force.

How many non-zero elements are necessary in $\langle A \rangle$? The number of non-zero elements is bound by $j$, simply as a consequence of basic mathematics.

Assume $|A_{\neq 0}| = n$. Such implies a bound of $O(2^{\log_2 n})$. The coefficients $(2i + 2)$ in the sum $\sum_{i=0}^{j}(2i + 2)a_i$ are even positive integer. Consequently any instance of $c = (2i + 2)$ can be expressed as a linear combination of any coefficients less than $c$. By induction it follows that $|A_{\neq 0}| \ll O(2^{\log_2 n})$.

 

Assume $b$ to be a constant independent of $n$ and $|A_{\neq 0}| \geq O((\log_2 n)^b)$. By the same argument of linear combination there must exist an $\langle A \rangle$ such that $|A_{\neq 0}| \ll O((\log_2 n)^b)$. If follows then that $|A_{\neq 0}| \leq O(\log_2 n)$ for some configuration of $\langle A \rangle$. The consequence of such is that the complexity of prime factoring reduces to the efficiency with which the non-zero elements of $\langle A \rangle$ can be determined.

One means to determine a configuration of $\langle A \rangle$ is to perform a binary search for inflection points determined by $\langle A \rangle$ in the interval $[0, j]$.

# 4 Figures

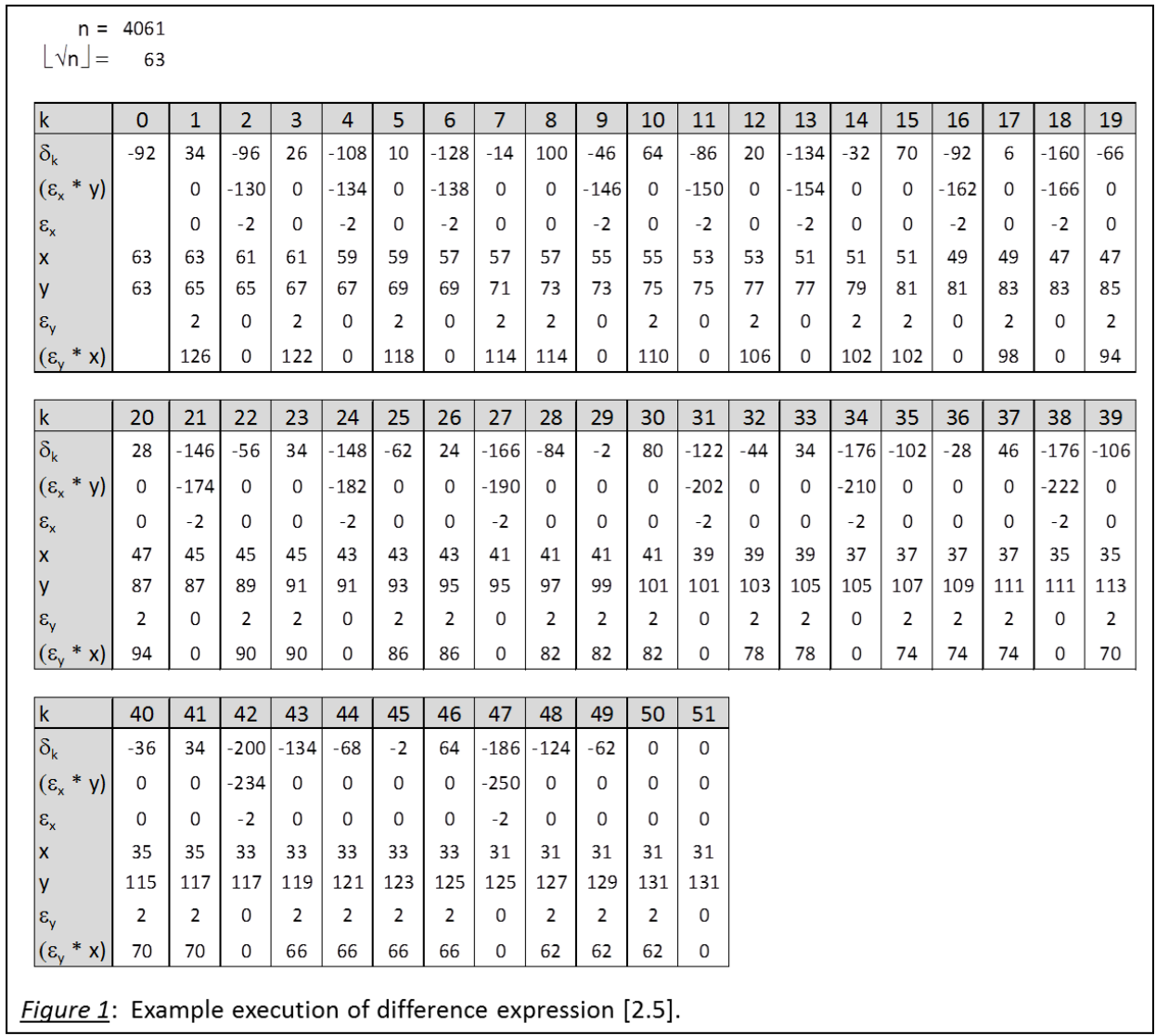

*Figure 1*: Example execution of difference expression [2.5].

    

n = 4061  
$\lfloor \sqrt{n} \rfloor = 63$

| k | 0 | 1 | 2 | 3 | 4 | 5 | 6 | 7 | 8 | 9 | 10 | 11 | 12 | 13 | 14 | 15 | 16 | 17 | 18 | 19 |
|---|---|---|---|---|---|---|---|---|---|---|---|---|---|---|---|---|---|---|---|---|
| $\delta_k$ | -92 | 34 | -96 | 26 | -108 | 10 | -128 | 100 | -46 | 64 | -86 | 20 | -134 | 70 | -92 | 6 | -160 | 28 | -146 | 34 |
| $(\varepsilon_x * y)$ |  | 0 | -130 | 0 | -134 | 0 | -138 | 0 | -146 | 0 | -150 | 0 | -154 | 0 | -162 | 0 | -166 | 0 | -174 | 0 |
| $\varepsilon_x$ |  | 0 | -2 | 0 | -2 | 0 | -2 | 0 | -2 | 0 | -2 | 0 | -2 | 0 | -2 | 0 | -2 | 0 | -2 | 0 |
| x | 63 | 63 | 61 | 61 | 59 | 59 | 57 | 57 | 55 | 55 | 53 | 53 | 51 | 51 | 49 | 49 | 47 | 47 | 45 | 45 |
| y | 63 | 65 | 65 | 67 | 67 | 69 | 69 | 73 | 73 | 75 | 75 | 77 | 77 | 81 | 81 | 83 | 83 | 87 | 87 | 91 |
| $\varepsilon_y$ |  | 2 | 0 | 2 | 0 | 2 | 0 | 4 | 0 | 2 | 0 | 2 | 0 | 4 | 0 | 2 | 0 | 4 | 0 | 4 |
| $(\varepsilon_y * x)$ |  | 126 | 0 | 122 | 0 | 118 | 0 | 228 | 0 | 110 | 0 | 106 | 0 | 204 | 0 | 98 | 0 | 188 | 0 | 180 |

| k | 20 | 21 | 22 | 23 | 24 | 25 | 26 | 27 | 28 | 29 | 30 | 31 | 32 | 33 | 34 |
|---|---|---|---|---|---|---|---|---|---|---|---|---|---|---|---|
| $\delta_k$ | -148 | 24 | -166 | 80 | -122 | 34 | -176 | 46 | -176 | 34 | -200 | 64 | -186 | 0 | 0 |
| $(\varepsilon_x * y)$ | -182 | 0 | -190 | 0 | -202 | 0 | -210 | 0 | -222 | 0 | -234 | 0 | -250 | 0 | 0 |
| $\varepsilon_x$ | -2 | 0 | -2 | 0 | -2 | 0 | -2 | 0 | -2 | 0 | -2 | 0 | -2 | 0 | 0 |
| x | 43 | 43 | 41 | 41 | 39 | 39 | 37 | 37 | 35 | 35 | 33 | 33 | 31 | 31 | 31 |
| y | 91 | 95 | 95 | 101 | 101 | 105 | 105 | 111 | 111 | 117 | 117 | 125 | 125 | 131 | 131 |
| $\varepsilon_y$ | 0 | 4 | 0 | 6 | 0 | 4 | 0 | 6 | 0 | 6 | 0 | 8 | 0 | 6 | 0 |
| $(\varepsilon_y * x)$ | 0 | 172 | 0 | 246 | 0 | 156 | 0 | 222 | 0 | 210 | 0 | 264 | 0 | 186 | 0 |

*Figure 2*: Example execution of difference expression [2.5] computing $\varepsilon_y$ to force $\delta_k \leq 0$.

n = 4061  
$\lfloor \sqrt{n} \rfloor = 63$

| k | 0 | 1 | 2 | 3 | 4 | 5 | 6 | 7 | 8 | 9 | 10 | 11 | 12 | 13 | 14 | 15 | 16 | 17 |
|---|---|---|---|---|---|---|---|---|---|---|---|---|---|---|---|---|---|---|
| $\delta_x$ | -92 | -96 | -108 | -128 | -46 | -86 | -134 | -92 | -160 | -146 | -148 | -166 | -122 | -176 | -176 | -200 | -186 |  |
| $(-2 * y_k)$ |  | -130 | -134 | -138 | -146 | -150 | -154 | -162 | -166 | -174 | -182 | -190 | -202 | -210 | -222 | -234 | -250 |  |
| x | 63 | 61 | 59 | 57 | 55 | 53 | 51 | 49 | 47 | 45 | 43 | 41 | 39 | 37 | 35 | 33 | 31 |  |
| y | 63 | 65 | 67 | 69 | 73 | 75 | 77 | 81 | 83 | 87 | 91 | 95 | 101 | 105 | 111 | 117 | 125 | 131 |
| $\varepsilon_k$ | 2 | 2 | 2 | 4 | 2 | 2 | 4 | 2 | 4 | 4 | 4 | 6 | 4 | 6 | 6 | 8 | 6 |  |
| $(\varepsilon_k * x_k)$ | 126 | 122 | 118 | 228 | 110 | 106 | 204 | 98 | 188 | 180 | 172 | 246 | 156 | 222 | 210 | 264 | 186 |  |
| $\delta_y$ | 34 | 26 | 10 | 100 | 64 | 20 | 70 | 6 | 28 | 34 | 24 | 80 | 34 | 46 | 34 | 64 | 0 |  |

*Figure 3*: Example execution of difference expression [2.6].

## 5 References

[1] D. M. Bressoud, Factorization and Primality Testing, Springer-Verlag, 1989.

[2] S. S. Wagstaff, Jr, The Joy of Factoring, American Mathematical Society, 2013.